\documentclass[]{spie}  

\usepackage{amsmath,amsfonts,amssymb}
\usepackage{graphicx}
\usepackage{aas_macros}
\usepackage[margin=2cm]{geometry}

\title{The inside-out, upside-down telescope: the Argus Array’s new pseudofocal design}

\author[a]{Nicholas Law}
\author[a]{Alan Vasquez Soto}
\author[a]{Hank Corbett}
\author[a]{Nathan Galliher}
\author[a]{Ramses Gonzalez}
\author[a]{Lawrence Machia}
\author[b]{Glenn Walters}
\affil[a]{Department of Physics and Astronomy, University of North Carolina at Chapel Hill, Chapel Hill, NC 27599-3255, USA}
\affil[b]{Department of Applied Physical Sciences, University of North Carolina at Chapel Hill, Chapel Hill, NC 27599-3255, USA}

\authorinfo{Further author information: N.M.L. E-mail:  nmlaw@physics.unc.edu}

\begin{document} 
\maketitle

\begin{abstract}
The Argus Optical Array will be the first all-sky, arcsecond-resolution, 5-m class telescope. The 55 GPix Array, currently being prototyped, will consist of 900 telescopes with 61 MPix very-low-noise CMOS detectors enabling sub-second cadences. Argus will observe every part of the northern sky for 6-12 hours per night, achieving a simultaneously high-cadence and deep-sky survey. The array will build a two-color, million-epoch movie, reaching dark-sky depths of $m_g$=19.6 each minute and $m_g$=23.6 each week over 47\% of the entire sky, enabling the most-sensitive-yet searches for high-speed transients, gravitational-wave counterparts, exoplanet microlensing events, and a host of other phenomena.  In this paper we present our newly-developed array arrangement, which mounts all telescopes into the inside of a hemispherical bowl (turning the original dome design inside-out). The telescopes' beams thus converge at a single ``pseudofocal'' point. When placed along the telescope's polar axis, this point does not move as the telescope tracks, allowing every telescope to simultaneously look through a single, unmoving window in a fixed enclosure. This telescope bowl is suspended from a simple free-swinging pivot (turning the usual telescope mounting support upside-down), with polar alignment afforded by the creation of a virtual polar axis defined by a second mounting pivot. This new design, currently being prototyped with the 38-telescope Argus Pathfinder, eliminates the need for a movable external dome and thus greatly reduces the cost and complexity of the full Argus Array. Coupled with careful software scope control and the use of existing software pipelines, the Argus Array could thus become one of the deepest and fastest sky surveys, within a midscale-level budget. 
\end{abstract}


\section{INTRODUCTION}
Time-domain sky surveys\cite{Rau2009, Law2009PASP, master_instrument, atlas_instrument, asassn_instrument, ztf_instrument, panstarrs_instrument, catalina_instrument, crts_instrument, des_program, goto_instrument} are generally designed to enable a range of cadences, between deep-drilling small fields for rapid transients, through to tiling the sky over multi-night periods. As large-scale detectors have become available, the sky coverage of individual telescopes has increased, allowing deep-drilling strategies to cover increasingly large sky areas in single shots (e.g. the TESS\cite{Ricker_2015} survey). Recently, the availability of mass-produced wide-field telescopes and new low-cost, low-noise, high-pixel-count sensors has enabled this strategy to be taken to the extreme: large telescope arrays that cover the entire sky in every exposure, co-adding those images over very long periods to achieve depth, and thus simultaneously achieving a high-cadence and deep survey over very large sky areas\cite{law-evryscope-2015, ratzloff-evryscope, law-argus-pasp}.

The Argus Array\cite{law-argus-pasp} is designed to cover an 8,000 square degree field of view with 1.4"/pixel sampling, a total of 55 GPix (Table 1). The array will consist of 900 individual telescopes, with 61 Mpix ultra-low-noise CMOS detectors capable of observing cadences as short as one second. With maximum data rates of 110GB/sec, only reduced portions of the data can be stored for offline analysis. The Argus-HDPS (Hierarchical Data Processing system; Corbett et al. 2022 [these proceedings]) is designed for realtime data analysis on an array of GPUs, with all-sky continuous sequences of hundreds or thousands of images allowing strong rejection of signals from satellite constellations such as Starlink. We are currently prototyping the Argus Array software and hardware in a series of systems. The largest, the 38-telescope Argus Pathfinder, will shortly begin full science operations.

Once constructed, however, a 900-telescope array will be among the most complex astronomical instruments yet built. A conventional design, an array of small groups of telescopes on many mounts, leads to thousands of exposed moving parts and optical components, and thus extreme maintenance requirements and costs compared to a monolithic telescope on the same large scale. Argus is designed to dramatically reduce operations and maintenance costs by mounting the entire telescope array inside a lab-like environment, reducing the number of moving parts to just a few, and protecting all the optics within a clean, thermally stable, sealed enclosure. Our original design for the Argus Array achieved this by mounting the telescopes looking out from a hemispherical dome\cite{law-argus-pasp}. Although this greatly reduced complexity and cost compared to a conventional telescope array by requiring only one tracking mount, it still required one external window per telescope, and the entire dome was required to move to track the sky. 

\begin{figure}
    \centering
    \includegraphics[width=0.8\columnwidth]{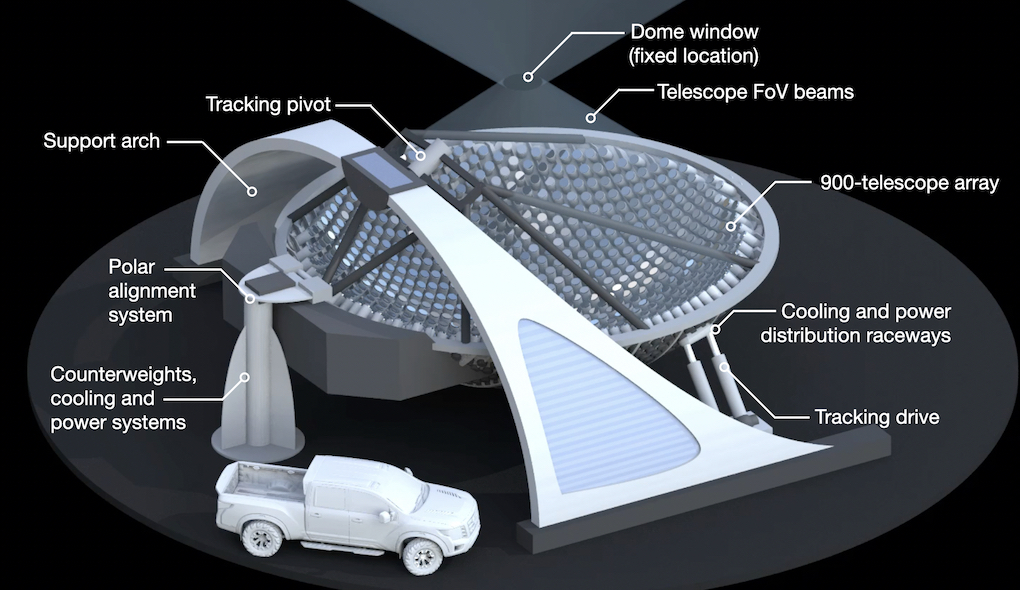}
    \caption{A conceptual overview of the Argus Array pseudofocal mechanical structure. The dome window is placed at the convergence point of the telescopes and does not require movement as the system tracks. A pickup truck is included for scale.}
    \label{fig:argus_concept_overview}
\end{figure}

\begin{figure}
    \centering
    \includegraphics[width=0.8\columnwidth]{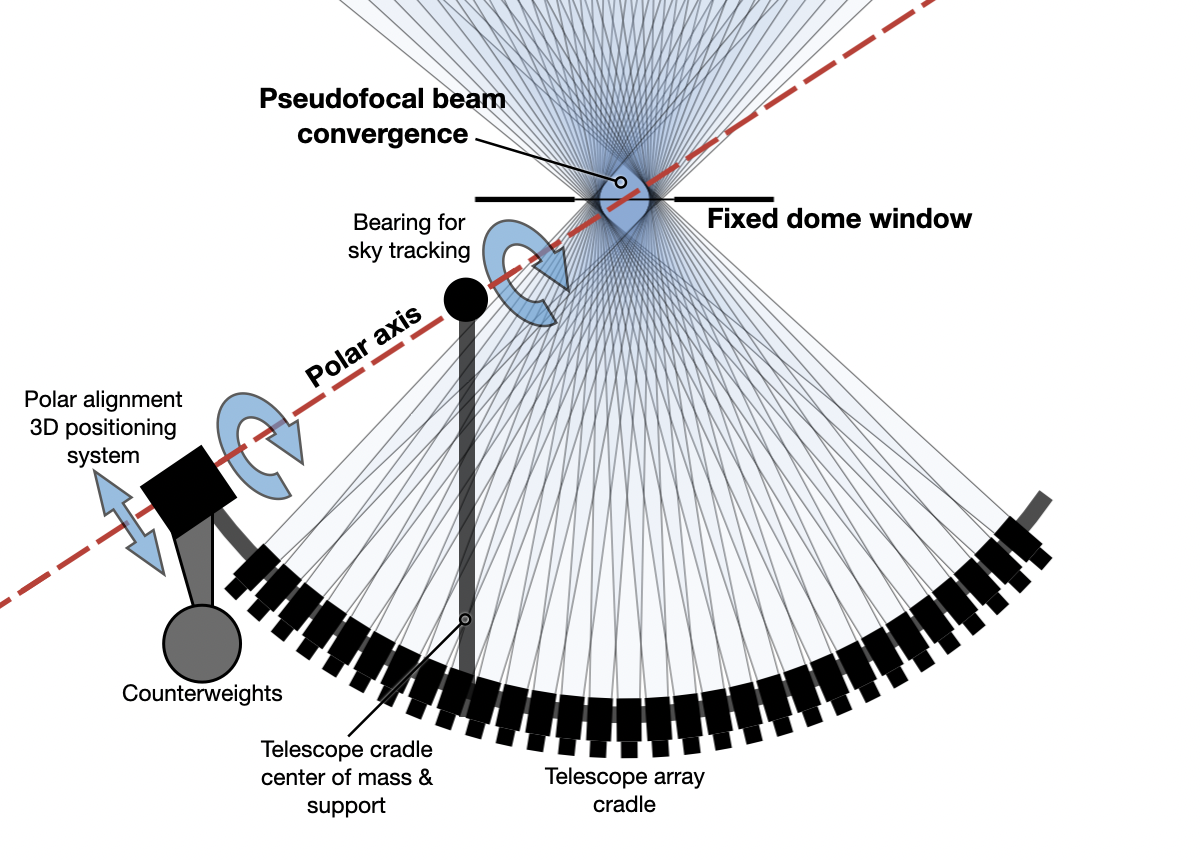}
    \caption{A simplified side-view of the main elements of a pseudofocal tracking array. The telescope beams converge at a window located along the polar axis of the system, which thus need not move during tracking. The system tracks the sky by a linear actuator pushing on the telescope cradle, which is constrained to rotate around the polar axis formed by the two system pivots. The upper pivot is freely rotating but fixed in position, with the telescope-cradle center of mass arranged directly beneath it. The lower pivot can be positioned in 3D space to change the polar axis direction and thus allow high-precision polar alignment.  }
    \label{fig:pseudofocal}
\end{figure}

During the Argus conceptual design phase, we developed an alternative ``inside-out'' concept, where the telescopes are mounted to the inside of a hemispherical bowl (Figure \ref{fig:argus_concept_overview}), causing a convergence of the telescope beams at the center of the hemisphere. This point looks somewhat like a focus in ray diagrams, as the light rays come together at that position. All beams entering the telescopes, however, are naturally collimated direct from the sky, and this point is just a optically-unimportant convergence we have dubbed the ``pseudofocus''. The convergence point is an area of physical space through which all the telescopes beams pass that can have an area two orders of magnitude smaller than the telescope array itself (Figures \ref{fig:argus_concept_overview}, \ref{fig:pseudofocal}). A single small window placed at that location in a fixed building can therefore protect all the telescopes simultaneously.

Furthermore, the system can be arranged such that this window (and thus the enclosure to which it is mounted) remains fixed as the array tracks the sky. Tracking, without causing image rotation, requires rotating the array around an axis parallel to the Earth's rotation axis. If we arrange the array geometry such that the window is along this rotation axis (Figure \ref{fig:pseudofocal}), the window can remain fixed in place during tracking (at least for the small tracking motions required for a full-sky array). By eliminating all external moving parts, this greatly simplifies the telescope enclosure compared to a conventional design: it becomes simply a building with a skylight.

The polar axis geometry affords an opportunity to track the entire telescope array by swinging it from two pivots along the  axis (also shown in Figure \ref{fig:pseudofocal}). This arrangement, upside-down compared to most telescope mounts, allows the large majority of the array weight to be supported by an off-the-shelf industrial ball-joint pivot, while the other much-smaller pivot maintains precision polar alignment.

In this paper we present an overview of the qualitative design considerations for a pseudofocal telescope array. The detailed Argus Array hardware and design is further explored in a series of papers in these proceedings. Corbett et al. 2022a details the rapid prototyping of the Argus core technologies in the Argus Array Technology Demonstrator (A2TD). The development of the Argus subsystems are covered in individual papers, including the A2TD mount (Gonzalez et al. 2022); the telescope array packing and field-of-view optimization (Galliher et al. 2022); the tracking and polar alignment system (Vasquez Soto et al. 2022); and the thermal and dome-seeing control system (Machia et al. 2022). Corbett et al. 2022b describes the Argus data processing pipelines and their development status.  In this paper, we discuss the geometry required to achieve a single fixed window (Section \ref{sec:array_geom}), tracking and precision polar alignment methods (Section \ref{sec:tracking}), optical considerations (Section \ref{sec:optics}), and the thermal design (Section \ref{sec:thermal}). We conclude with a discussion of the series of Argus Array prototypes designed to test the pseudofocal design (Section \ref{sec:concs}).

\section{PSEUDOFOCAL ARRAY DESIGN}

There are four major considerations in a fully-developed pseudofocal array design: a) arranging the telescope array geometry such that all telescopes can look though a single small window; b) arranging the telescope mount geometry such that the window remains fixed as the telescope tracks; c) designing the system optics to minimize ghosts and optical aberrations; and d) designing the system enclosure to eliminate turbulence and air motion around the array.

\subsection{Array geometry}
\label{sec:array_geom}
Because the beams must all pass through the same point, the physical telescope array shape mirrors the shape of the array’s field of view (FoV) on the sky. Since the sky forms a sphere, the simplest array shape is thus a bowl. If the array is designed to be gapless, the radius-of-curvature of that bowl ($R_c$) is set by the field of view of the telescopes ($F$) and the telescope aperture size ($d_{ap}$). A practical array, however, will need small physical gaps ($d_{ap}$) between telescopes, allowing for mounting and alignment systems. For a spherical telescope bowl, these quantities are  related in the following way:

\begin{equation}
    2\pi R_c \left(\frac{F}{360^\circ}\right) = d_{ap} + d_{gap}
\end{equation}

For the Argus Array, this leads to a radius of curvature of $\approx$6m, and thus a telescope bowl diameter of $\approx$10m (the bowl diameter is less than the base sphere's diameter because the telescopes do not cover a full $180^\circ$ FoV). This simple relationship assumes  one-dimensional linear packing; in a real design (Galliher et al. 2022, these proceedings) more complex 2D packing arrangements allow somewhat more compact arrays.

\subsection{Tracking system \& fixed window geometry}
\label{sec:tracking}
Precision sky tracking, without the image rotation inherent to an altitude/azimuth mount, requires the entire telescope array to rotate around an axis aligned to the Earth's rotation axis. The array’s polar axis is a line that need only be defined by two points. If we hang the array from those points, allowing free rotation around them, the full array is thus constrained to rotate only around the axis defined by those points. Polar alignment can then be easily achieved by constructing a single, fixed-in-place pivot from which the array hangs, and adding a second pivot which can be precisely located in 3D space to fix the direction of the polar axis. 

We have elected to place the array center of mass directly below the fixed pivot. With potentially tens of thousands of pounds to support, this allows the load-bearing pivot to be a simple off-the-shelf industrial item carrying the vast majority of the structural load. The precision-polar-alignment pivot, then, need only take on lateral load as the system tracks, at most a few percent of the total weight. Both the pivots and the window must be located along the polar axis, and symmetry requires that the array's natural center of mass is close to its optical convergence point. In practice, therefore, counterweights are required to move the center of mass away from the window location. 

To actually move the system, a linear actuator pushes the telescope bowl from the side. Because the linear actuator is at a long lever arm compared to the pivot position, the effects of motion imperfections are minimized, and our current prototypes have shown that off-the-shelf components achieve the required tracking accuracy (Vasquez Soto et al. 2022, these proceedings). We expect the array to track for 15 minutes (approximately one camera field of view in sky motion) at a time, and so the tracking motion is only several degrees.

Although the telescope array is on the scale of a large optical telescope, the Argus tracking system is designed for only the low-amplitude swings the all-sky survey requires, and is built from almost entirely off-the-shelf industrial components. The tracking drive can thus be constructed at, relatively, a very low cost.

\subsection{Optical considerations}
\label{sec:optics}
The pseudofocal array uncommonly requires all light to pass through an external glass window which appears quite tilted for almost all telescopes in the array. Tilted optics are common in internal telescope instrumentation (for example in dichroic beamsplitters), and do not introduce optical aberrations. In converging beams they can, however, induce ghost images from double-reflections, beam displacement, and other undesirable effects. In this subsection we show that almost all of these negative effects are not present in the pseudofocal design (because only parallel beams pass through the window), and discuss mitigation strategies for the remaining effects.

\subsubsection{Enclosure window optics}
\noindent{\bf Window shape and size:} The dimensions of the fixed window are defined by the footprint of the convergence point of all the telescope beams. A single telescope looking through the window with a beam direction normal to the window has a beam size of the telescope aperture ($D_{ap}$) plus the beam expansion angle corresponding to the telescope’s field of view ($F$). Because the physical beam expansion increases with distance between the telescope and window ($R_C)$, the footprint size also depends on the radius of curvature of the telescope bowl and thus on the mechanical packing efficiency of the telescopes. Telescopes looking though the window at more oblique angles ($\theta$) also have an elongated beam footprint. For the full array, the window diameter  $D_W$  is determined by the beam size of a telescope at the edge of the array:
\begin{equation}
D_W = \frac{2R_C\tan(F/2) + D_{ap}}{cos(\theta)}
\end{equation}

This relation shows that an increase in the number of telescopes, overlapping the telescope FoVs, increasing the telescope aperture, etc., correspondingly requires an increase in window size. The Argus bowl design maximizes the telescope-to-window ratio for a given full-array FoV, leading to a 36-inch window, $\approx$4$\times$ larger than the individual telescope apertures, but $<1\%$ of the surface area of the full telescope array.

\noindent{\bf Angle-of-incidence effects:} Commercially-available antireflection coatings have a restricted range of effective angles of incidence, and this places an effective upper limit of $100^{\circ}$--$120^{\circ}$ on the field of view of an array looking though a single window. The Argus Array's Key Projects' requirements\cite{law-argus-pasp} easily conform to these limits, especially with the acceptance of slightly increased light losses for telescopes at the edge of the array. Newly-developed coatings\cite{ar_coating} could relax these limits in the future. We have also explored multiple-window designs, but found that the increased mechanical complexity outweighed the optical performance gains.

\noindent{\bf Window optical quality:} Parallelism of the window glass is critically important to avoid imparting a poor wavefront onto the telescopes. However, a meniscus-lens-like bending of the window (the most common concern for a necessarily-unsupported horizontal piece of glass) does not significantly affect the final image quality.  The relatively small size of the window ($<$90cm even for the full 900-telescope Argus Array) thus makes parallelism at the required level relatively simple to achieve. Our Zemax and on-sky experiments for the Argus prototypes have confirmed that even high-quality float glass produces optical quality within the error budget for acceptable images across the array.

\subsubsection{Optical ghosts}
\label{sec:ghost_removal}
\noindent{\bf Internal window ghosts:} The telescope window passes only naturally collimated beams from the sky, and so this tilt cannot induce aberrations (for well-figured and parallel glass surfaces). Beam displacement is induced, but this has no effect on the telescope images. Similarly, multiple reflections within the window cannot produce ghosts because multiple-bounce reflections of parallel beams only displace the beam positions.

\noindent{\bf Inter-telescope ghosts:} The pseudofocal design introduces one extra, and complex, source of ghosts in the array images: light reflected back from the telescopes, off the fixed dome window, and into another telescope in the array. Because they involve a bounce out though the entire telescope optical train, these reflections can produce focused stellar images on other telescopes in the array; furthermore, the beams are reflected around the normal axis of the window and so ghosts can be induced between widely-separated telescopes. Because the window is fixed while the telescope tracks, the ghosts would move rapidly through the field, potentially producing confusing sources for transient detection.

\begin{figure}
    \centering
    \includegraphics[width=0.8\columnwidth]{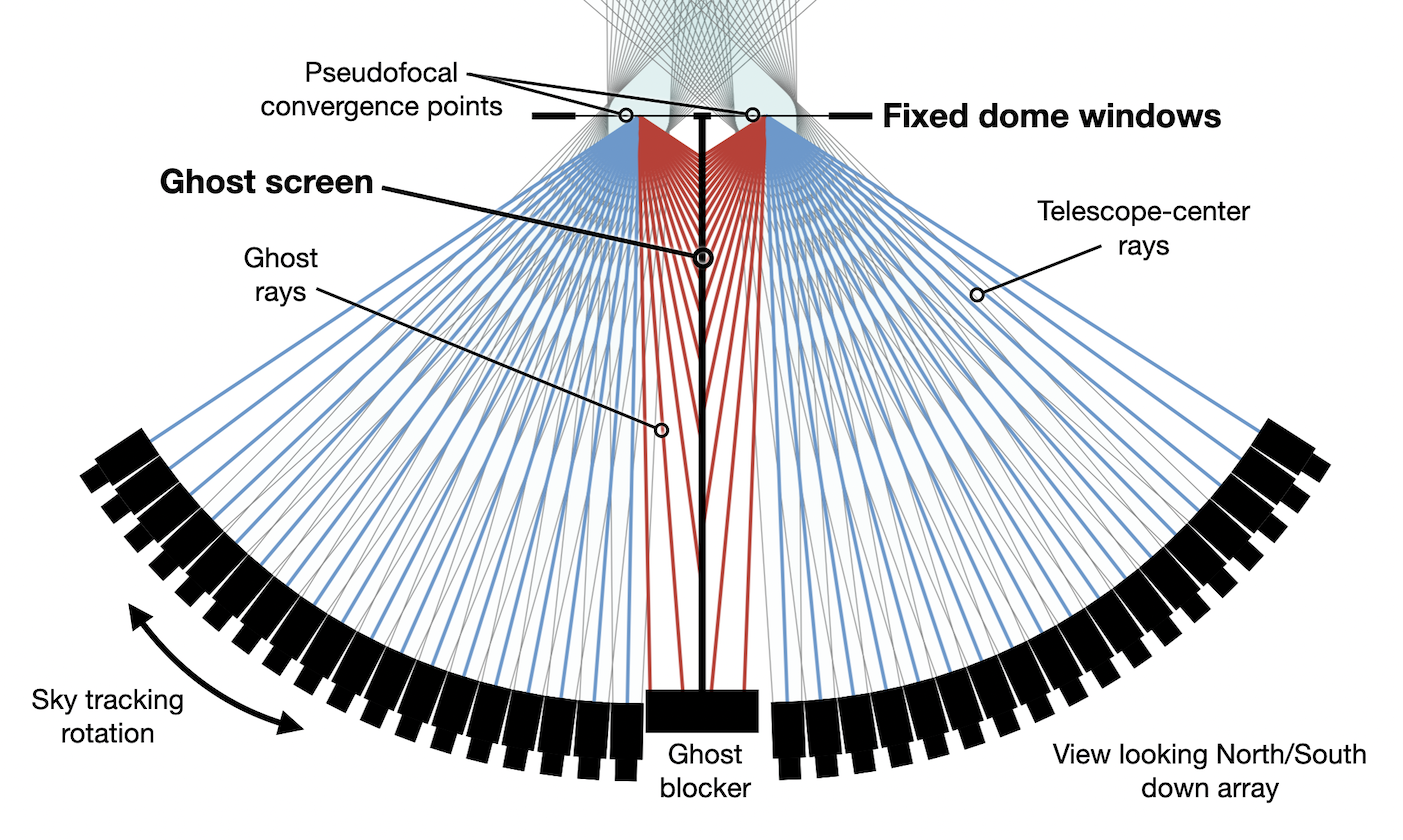}
    \caption{Blocking double-reflected ghosts with a two-window split array. The entire array is still mounted on a single tracking mount, but a dividing wall prevents all two-bounce optical ghosting between telescopes.}
    \label{fig:ghost_blocker}
\end{figure}

This issue can be mitigated by high-quality antireflection coatings, and that is likely to be sufficient for most applications. Furthermore, there is a simple mechanical solution that completely interrupts all inter-telescope beam paths. The system geometry requires that all reflected ghost beams cross the polar axis, producing optical ghost images on the other side of the array. A simple screen between the east and west sides of the array thus blocks all possible inter-telescope ghosts produced by the pseudofocal design. A small split between the east and west halves of the array, with each looking though their own window, enables this ghost screen to avoid all science photons (Figure \ref{fig:ghost_blocker}). The entire array is still mounted on a single (slightly larger) mount, and thus this solution thus does not add significant cost.

\subsubsection{Scattered light}
With a tiny window opening making up only $\approx0.2\%$ of the dome surface area, the pseudofocal array is  less susceptible to scattered light than a conventional dome design. Each telescope is effectively looking through a distant stop, and moonlight (for example) can thus only enter the dome as a narrowly-constrained beam. Scattered-light control is thus reduced to two areas:

\noindent{\bf Window dust accumulation:} As the only optical surface exposed to unfiltered air, the fixed window is susceptible to collecting dust, pollen and other contaminants. We have designed the window as a weather-sealed unit (although the sunshield gives further protection during the day), and its antireflection coatings are designed for architectural glass. The window can thus be regularly cleaned in the same manner as a building window, with a pressure washer or squeegee. A baffle around the window restricts light entry to the array's field of view.

\noindent{\bf Internal scattered light:} Moonlight and starlight which makes it into the dome will have an approximately 50\% chance of hitting a telescope and sensor as science-photons. The remainder, falling into mechanically-required gaps in the array, can be controlled by optically-absorbing surfaces, black-painted or black-anodized components, and light shrouds around the sensors themselves.

\begin{figure}
    \centering
    \includegraphics[width=0.8\columnwidth]{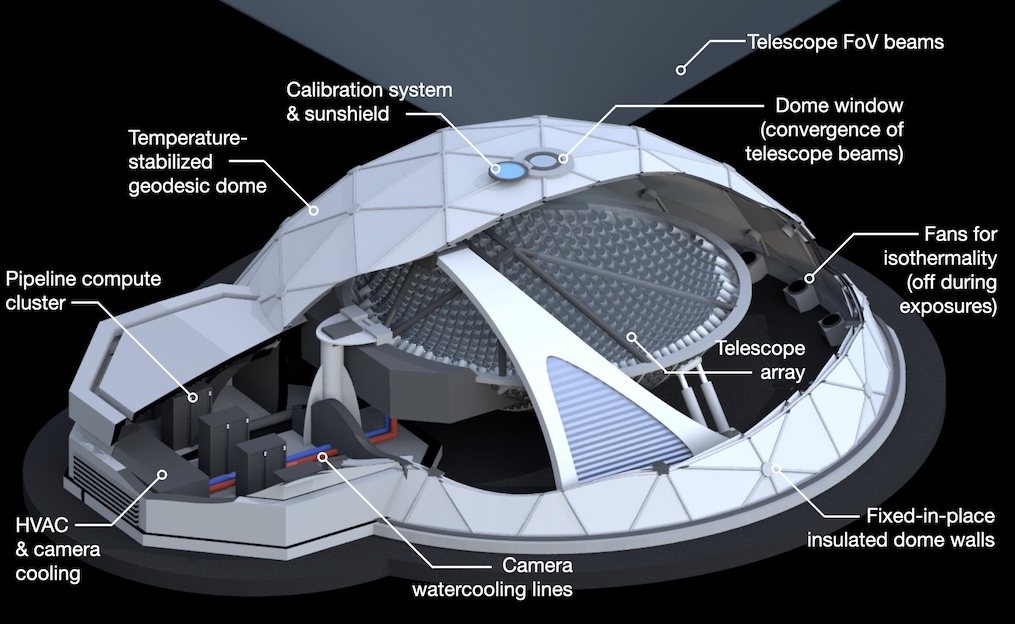}
    \caption{The Argus Array within its enclosure. The fixed enclosure is constructed from a low-cost, heavily-insulated geodesic dome with an extension housing the computing an HVAC systems. Heat generation within the enclosure is minimized by water cooling all electronic components, and fans are used between exposure runs to maintain an isothermal environment.}
    \label{fig:enclosure}
\end{figure}

\subsection{Thermal control and enclosure}
\label{sec:thermal}
With a single fixed window at the top of the dome, careful thermal control is essential to avoid dome seeing being generated. Without any openings, traditional dome seeing reduction is not possible. Instead, the array enclosure (Figure \ref{fig:enclosure}) will be designed to maintain isothermality as far as possible, preventing the generation of convective cells. This protection will be maintained by a series of passive and active measures. 
 
\subsubsection{Decoupling the array from the external thermal environment}
The Argus dome will be heavily insulated, with an internal temperature maintained over long (seasonal) periods, and an external temperature close to that of the surrounding environment (to avoid telescope-induced ground seeing). The Argus Pathfinder prototype will thoroughly test the ability of passive insulation to maintain the enclosure external at a temperature sufficiently close to that of the surrounding air. If necessary, an active system in a double-shelled design would enable the interior and exterior surfaces to be essentially decoupled, eliminating the risk of dome seeing or warm-dome-induced external seeing. A sunshield over the window will protect the telescopes during the daytime, as well as provide a flat-field calibration system.

The window is the only surface for which temperature maintenance could be a challenge. For Argus, the window only makes up $\sim0.1\%$ of the surface area of the enclosure. The telescopes, however, are imaging through the window, and so we must avoid strong turbulence being generated at the window/air interface.  The window will be thick (1/4 inch or more) to maintain appropriate flatness and weather resistance, and the internal enclosure temperature will be changed seasonally to avoid large temperature gradients. The Argus Pathfinder will verify this approach in realistic mountaintop conditions and at almost full scale (because beam elongation dominates the window size, the Pathfinder window is $\approx$75\% of the full Argus Array's window size). If further insulation is necessary, a double-glazing window scheme could be built, with resulting ghosts controlled by the design explored in Section \ref{sec:ghost_removal}.

\subsubsection{Eliminating inside-enclosure heat sources \& ensuring an isothermal environment}
All heat-producing equipment inside the dome (most notably, the science cameras) will be cooled by an insulated watercooling system which removes generated heat from the enclosure without causing significant temperature gradients. Other electrical components such as network switches are located within cabinets watercooled by the same system.  This, along with the dome insulation, greatly limits hot-spots and gradients that could lead to convective turnover. 

Even a very heavily insulated dome will, however, still transport heat through its surface and may eventually cause thermal gradients to form between the array and the enclosure walls. Airflow for internal heat conduction could prevent these issues, but at the potential cost of introducing turbulence. The system tracking design, however, requires a $\approx$1 minute downtime to reset the tracking drive every 15 minutes. No observations take place during this return-to-start motion, and this period can thus be used for aggressive thermal control measures. Fans fed with filtered air at the required enclosure temperature will ensure the internal air is well-mixed, and give the opportunity for rapid thermal transfer with any array components that need to be returned to the correct temperature. With the fans turned off just before the start of exposures, and heat sources within the dome carefully controlled, we expect the settling time to be seconds, leaving an isothermal environment with little opportunity for dome seeing to arise. The Pathfinder telescope is designed with a 1/2.5-scale version of this system and will validate its performance.

\section{THE ARGUS PROTOTYPE SERIES \& SUMMARY}

\begin{table}[b]
    \centering
    \includegraphics[trim={3.7cm 15.3cm 27cm 4cm},clip,width=1.0\columnwidth]{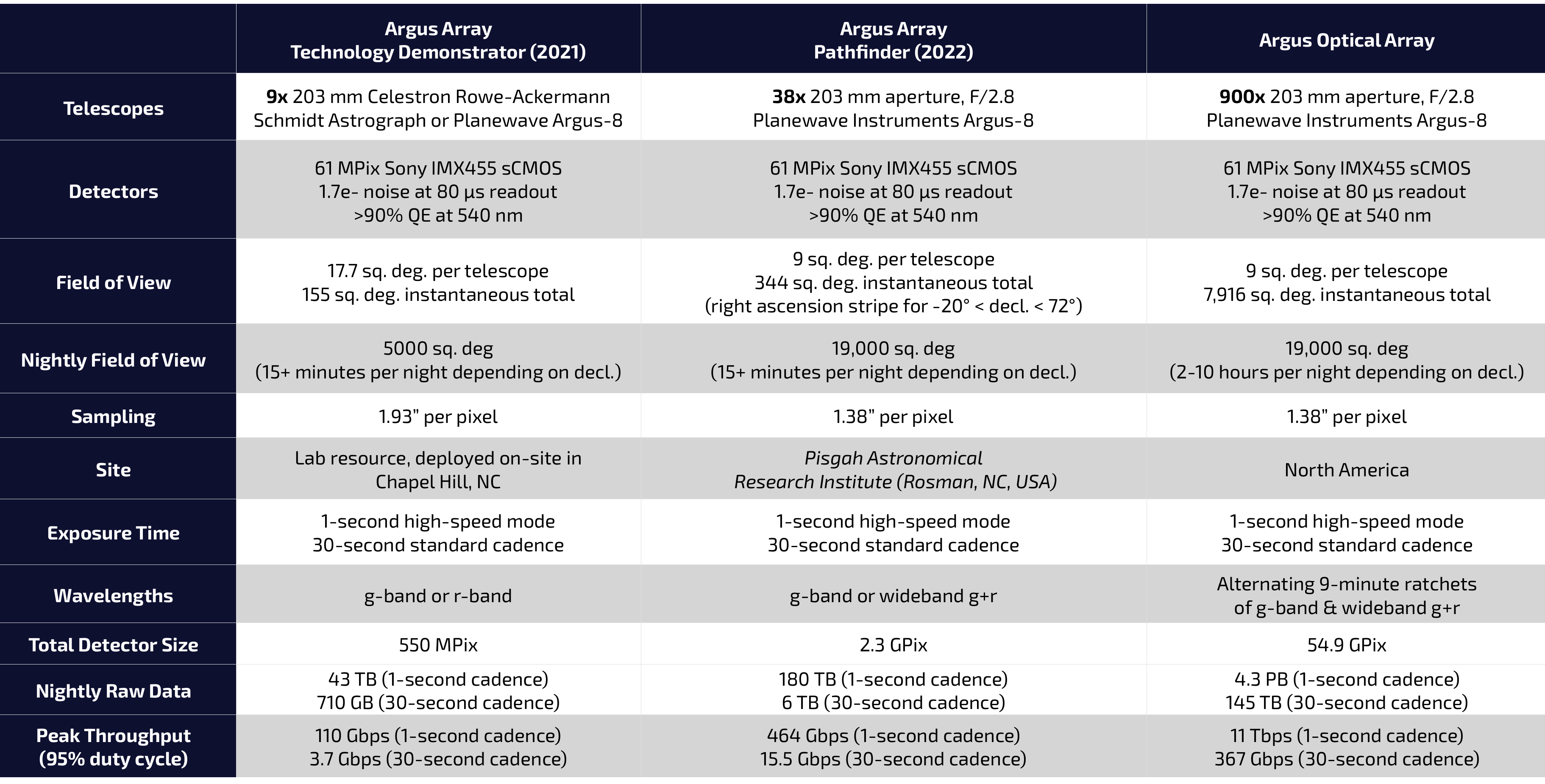}
    \caption{The specifications of the Argus Array prototypes compared to the full array baseline design.}
    \label{fig:my_label}
\end{table}

\label{sec:concs}
We are testing the Argus Array design in a series of prototypes (Table 1). The Argus Technology Demonstrator went on-sky in 2021 and is used for the development of the telescopes, cameras and tracking drive systems. The 38-telescope Argus Pathfinder is currently under construction and is a 1/2.5 scaling in linear dimension of cradle compared to the full array. Pathfinder will validate the pseudofocal design's optical, mechanical and thermal performance with a telescope ``cradle'' that forms a segment of the full Argus Array's bowl. Pathfinder will also demonstrate the operation of the full Argus pipeline, including rapid image astrometry, de-warping, segmentation, and image subtraction. Starting in 2022, Pathfinder will perform a sweep through the sky each night, with 15 minutes of high-cadence observations on each part of the sky over a $92^\circ$ declination range. This dataset, which will be publicly released, will enable a range of initial Argus key projects\cite{law-argus-pasp} including fast-radio-burst optical counterparts, fast nearby-star flare followup, and solar-system occultations. 

\acknowledgments 
This paper was supported by NSF MSIP (AST-2034381) and a grant from Schmidt Futures. This research, and the construction of the Argus prototypes, is undertaken with the collaboration of the Be A Maker (BeAM) network of makerspaces at UNC Chapel Hill and the UNC BeAM Design Center.

\bibliography{argus} 
\bibliographystyle{spiebib} 

\end{document}